# Integrated photonics incorporating 2D materials for practical applications

David J. Moss

Optical Sciences Centre, Swinburne University of Technology, Hawthorn, VIC 3122, Australia

Email: dmoss@swin.edu.au

## ABSTRACT

On-chip integration of 2D materials with exceptional optical properties provides an attractive solution for next-generation photonic integrated circuits to address the limitations of conventional bulk integrated platforms. Over the past two decades, significant advancements have been made in the interdisciplinary field of 2D material integrated photonics, greatly narrowing the gap between laboratory research and industrial applications. In this paper, we provide a perspective on the developments of this field towards industrial manufacturing and commercialization. First, we review recent progress towards commercialization. Next, we provide an overview of cutting-edge fabrication techniques, which are categorized into large-scale integration, precise patterning, dynamic tuning, and device packaging. Both the advantages and limitations of these techniques are discussed in relation to industrial manufacturing. Finally, we highlight some important issues related to commercialization, including fabrication standards, recycling, service life, and environmental implications.

## I. INTRODUCTION

Integrated circuits (ICs) have been the driving force behind the information age. Since the first demonstration of a prototype IC in 1958 by Nobel laureate Jack Kilby,[1] many technological breakthroughs have rapidly advanced their development, continually improving their core benefits including compact footprint, low power consumption, high scalability, and cost-effective mass production. After more than 60 years of development, ICs have now become a cornerstone of modern industry, driving revolutionary changes worldwide and profoundly impacting people's lives.

In modern IC industry, the complementary metal-oxide-semiconductor (CMOS) fabrication is a dominant manufacturing technology with well-established production lines and fabrication standards.[2,3] By leveraging the well-developed CMOS fabrication technology, photonic integrated circuits (PICs) not only reap the dividends of electronic ICs with respect to the device footprint, energy consumption, and mass producibility, but also provide additional benefits such as large processing bandwidth, strong immunity to electromagnetic interference, and capability for massively parallel processing.[4-6] These make them competitive for overcoming the intrinsic bandwidth bottleneck of electronic ICs in many high-bandwidth applications.[7,8]

As the CMOS technology continues to push the boundaries of device miniaturization, the inherent limitations in material properties of conventional PICs pose challenges in meeting the ever-increasing demands for device functionality and performance.[9] For example, as a dominant integrated material platform, silicon's indirect bandgap hinders its use in laser applications.[10] In addition, the strong two-photon absorption in silicon at the near-infrared wavelengths greatly limits its effectiveness in many nonlinear optical applications.[11]

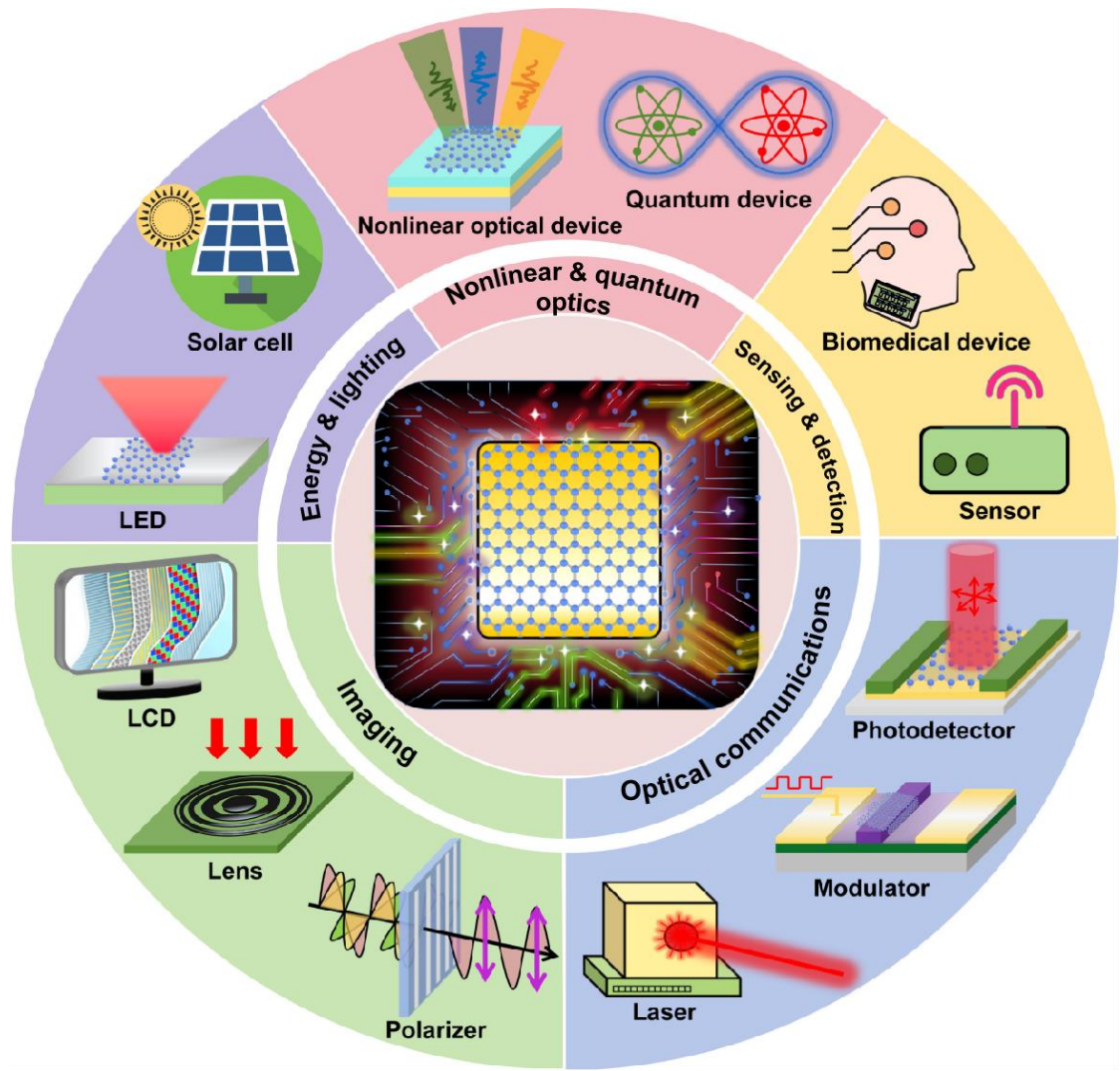

**FIG.1.** Diverse applications of integrated photonic devices incorporating 2D materials. LCD: liquid crystal display. LED: light emitting diode

On-chip integration of other functional materials with superior properties has proven to be a promising solution to address the limitations of conventional PICs. Since the first isolation of graphene in 2004,[12] atomically thin two-dimensional (2D) materials have received enormous interests, with a fast-growing family including transition metal dichalcogenides (TMDCs), graphene oxide (GO), black phosphorus (BP), hexagonal boron nitride (hBN), and many others. Unlike their bulk counterparts, 2D materials demonstrate a range of exceptional properties, such as ultrahigh carrier mobility, broadband optical response, layer-dependent optical bandgaps, significant material anisotropy, and high optical nonlinearity.[13-18] Moreover, unlike the common lattice mismatch issue faced with bulk integrated materials, 2D materials are feature by dangling-bond free surface and an inherent propensity to adhered through van der

Waals.[19,20] This allows for stress-free on-chip integration and provides a high flexibility in device design.

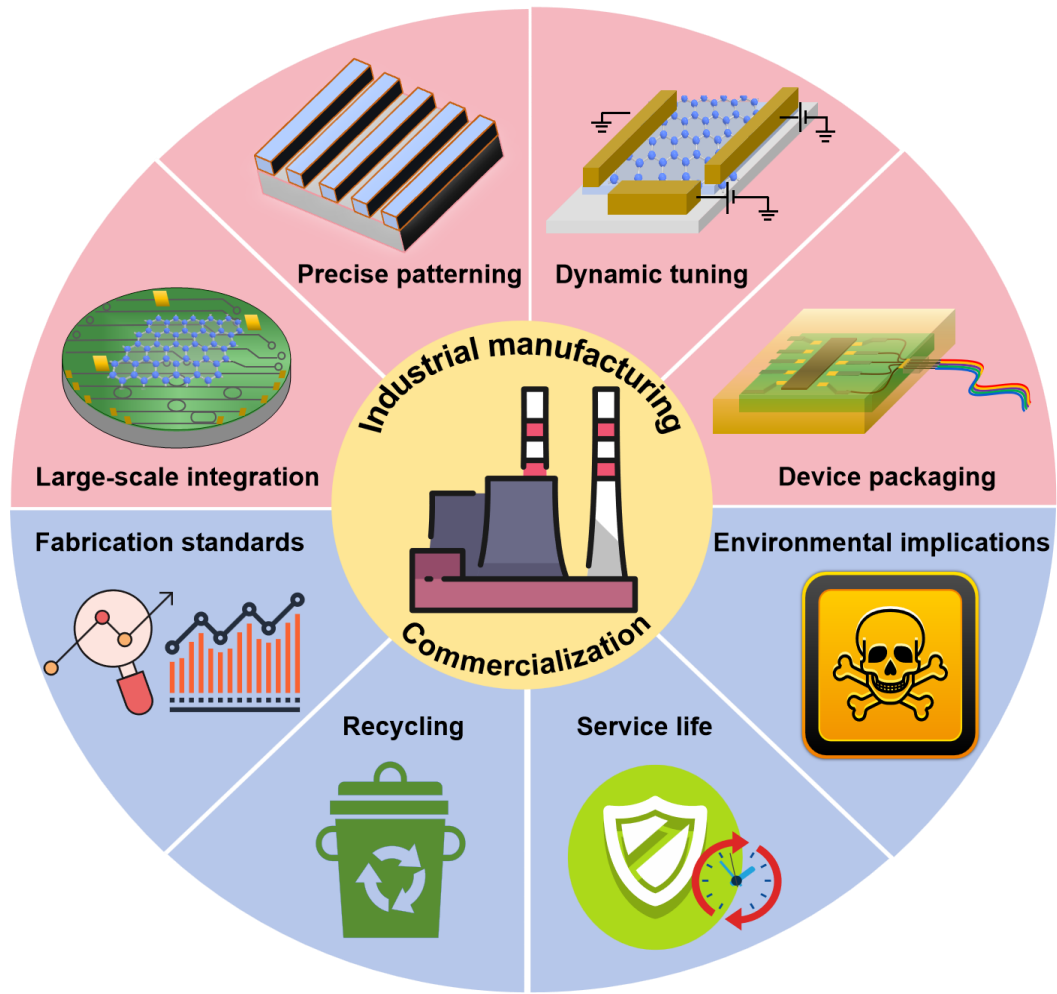

**FIG.2.** Key requirements for industrial manufacturing and commercialization of integrated photonic devices incorporating 2D materials.

Over the past two decades, significant progress has been made in the interdisciplinary field of 2D material integrated photonics. A number of integrated photonic devices incorporating 2D materials have demonstrated new functionalities and improved performance that surpass what the conventional bulk integrated photonic devices can offer,[21-28] covering a wide range of applications as summarized in **Fig. 1**. The huge progress in 2D material integrated photonics is not only intriguing for laboratory research but also paves the way for many industrial applications.

In this paper, we provide a perspective on the development of 2D material integrated photonics, with a particular focus on its trajectory towards industrial manufacturing and commercialization. We review the current progress, discuss future opportunities, and highlight the remaining gap between the two. The structure of this perspective is illustrated in **Fig. 2**. First, we review recent progress towards commercialization of integrated photonic devices incorporating 2D materials. Next, we provide an overview of cutting-edge fabrication techniques, which are categorized into large-scale integration, precise patterning, dynamic tuning, and device packaging. We discuss both their merits and limitations in the context of industrial manufacturing. Finally, we discuss some critical issues related to commercialization, including fabrication standards, recycling, service life, and environmental implications.

## II. COMMERCIALIZATION PROGRESS

Since the ground-breaking work on isolation of graphene in 2004, 2D materials have become a highly active research area over the past two decades, as evidenced by a wealth of related publications and patents. Along with extensive research on these atomically thin materials, there has also been growing interest in the pursuit of 2D materials in commercial products. In the past decade, exciting progress has been made in bridging laboratory research to industrial manufacturing. This section provides an overview of these advancements in integrated photonic devices incorporating 2D materials.

**Fig. 3** shows some examples of state-of-the-art hybrid integrated photonic devices with 2D materials, including both commercially available products and laboratory-successful devices with a high level of market readiness. **Fig. 3(a)** shows a 10-Gb/s integrated graphene phase modulator demonstrated by an international team led by the Photonic Networks and Technologies National Laboratory in Italy.[29] By integrating the graphene phase modulator into a Mach-Zehnder interferometer configuration, a static modulation depth of ~35 dB and a modulation efficiency of ~0.28 V·cm was achieved, which outperform silicon-based PN junctions and comparable to modulators based on silicon-insulator-silicon capacitors. **Fig. 3(b)** shows a commercially available graphene photodetector from the Emberion corporation.[30] By integrating graphene and nanostructured optical absorbers with specially designed read-out circuits, it was capable of detecting light across a wide wavelength range of ~400 – 1800 nm. **Fig. 3(c)** shows a graphene optoelectronic mixer demonstrated by the Thales Research and Technology in collaboration with the University of Lille and IEMN and the University of Cambridge.[31] It achieved an operation bandwidth of ~20 GHz, with a conversion efficiency surpassing those of previous graphene mixers by more than two orders of magnitude.

**Fig. 3(d)** shows a high-efficient perovskite solar cell designed by the Swift Solar corporation.[32] By combining metal halide perovskites with silicon substrates, this tandem solar cell was able to surpass the 30% conversion efficiency barrier that limited traditional solar cells. **Fig. 3(e)** shows a graphene-GaN light-emitting diode (LED) demonstrated by the Gwangju Institute of Science and Technology in Korea.[33] It achieved a sheet resistance of ~620 ohms per square and a transparency exceeding 85% in the 400 – 800 nm wavelength range, outperforming conventional GaN LEDs with indium tin oxide. **Fig. 3(f)** shows GO hologram demonstrated by researchers from the University of Shanghai for Science and Technology.[34] The hologram image was patterned via direct laser writing and was able to be erased via oxygen plasma treatment, providing a facile, low-cost, and rewritable method for hologram imaging.

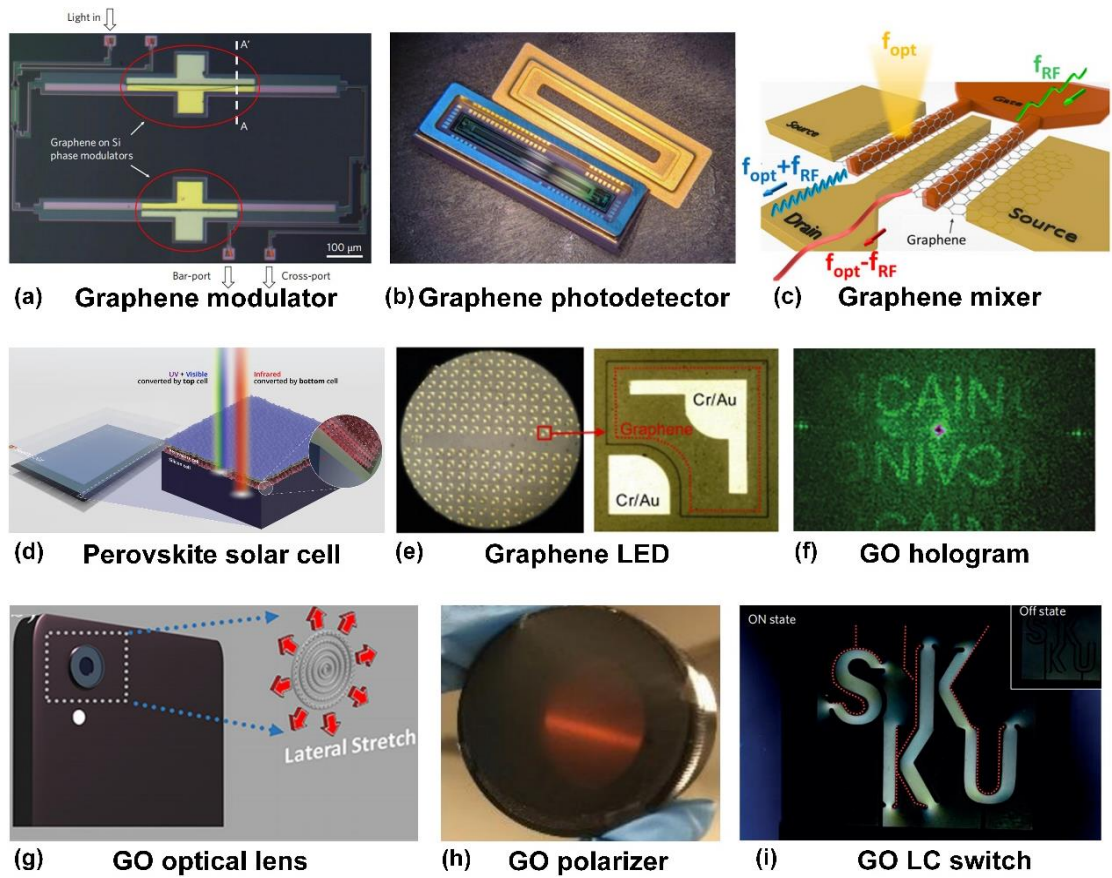

**FIG. 3.** Commercialization of integrated photonic devices incorporating 2D materials. (a) A graphene phase modulator from the Photonic Networks and Technologies National Laboratory in Italy. (b) A graphene photodetector from the Emberion corporation. (c) A graphene optoelectronic mixer demonstrated by the Thales Research and Technology in collaboration with the University of Lille and IEMN and the University of Cambridge. (d) A perovskite solar cell from the Swift Solar corporation. (e) A graphene-GaN light-emitting diode (LED) demonstrated by researchers from the Gwangju Institute of Science and Technology. (f) GO hologram demonstrated by researchers from the University of Shanghai for Science and Technology. (g) A GO optical lens demonstrated by researchers from Shenzhen University and Swinburne University of Technology. (h) A GO optical polarizer demonstrated by researchers from Swinburne University of Technology. (i) A GO liquid-crystal (LC) switch demonstrated by researchers from Sungkyunkwan University. (a) Reproduced with permission from Sorianello *et al.*, Nat. Photonics **12**(1), 40-44 (2017). Copyright 2017 Springer Nature. (b) Reproduced from https://optics.org/news/10/6/18. (c) Reproduced with permission from Montanaro *et al.*, Nat. Commun. **12**(1), 2728 (2021). Copyright 2021 Springer Nature. (d) Reproduced from https://www.swiftsolar.com/tech/. (e) Reproduced with permission from Jo *et al.*, Nanotechnology **21**(17), 175201 (2010). Copyright 2010 IOP Publishing. (f) Reproduced with permission from Wan *et al.*, Adv. Opt. Mater. **11**(22), 2300872 (2023). Copyright 2023 John Wiley and Sons. (g) Reproduced with permission from Wei *et al.*, ACS Nano **15**(3), 4769-4776 (2021). Copyright 2021 American Chemical Society. (h) Reproduced with permission from Zheng *et al.*, Nanoscale **12**(21), 11480-11488 (2020). Copyright 2020 RSC Pub. (i) Reproduced with permission from Shen *et al.*, Nat. Mater. **13**(4), 394-9 (2014). Copyright 2014 Springer Nature.

**Fig. 3(g)** shows a GO optical lens demonstrated by researchers from Shenzhen University and Swinburne University of Technology.[35] The ultrathin lens, with a thickness of ~250 nm, enabled color imaging and simultaneous focal length tunability across the entire visible spectrum. By laterally stretching the lens, it achieved over 20% focal length tuning, surpassing traditional lenses that rely on tuning the distance between multiple optical elements. **Fig. 3(h)** shows a GO optical polarizer (mounted on a commercial standard polarizer mount) demonstrated by researchers from Swinburne University of Technology.[36] This device achieved a high polarization extinction ratio of ~ 20 dB, along with adjustable operation wavelengths in the mid-infrared region. **Fig. 3(i)** shows a GO-liquid crystal (LC) switch demonstrated by Sungkyunkwan University,[37] achieving 15:1 contrast between the on and off states under an applied voltage of ~20 V. The large polarizability anisotropy of GO also yielded a large Kerr coefficient that was about three orders of magnitude higher than that for molecular LCs.

Now the research on 2D materials is at a stage where most of outstanding properties of conventional 2D materials such as graphene, GO, TMDC, hBN have already been studied, and the focus is gradually shifting towards industrial-scale production and the development of commercial products for these materials. In the past decade, there has been a surge of related commercial activities, with efforts focused on bringing relevant products to market. For instance, the Graphene Flagship initiative, which brings together 118 academic and industrial partners, launched projects such as METROGRAPH, GBIRCAM, AUTOVISION, and GRAPES to explore graphene-based devices for displays, imaging technologies, and solar cells.[38] The INNOFOCUS corporation seeks to leverage 2D materials with innovative nanostructures to develop advanced photonic devices, catering to industrial requirements across various applications.[39] On the other hand, studies on a range of new 2D materials, such as MXenes, metal-organic frameworks (MOFs), and ferroelectric materials, continue to emerge and reveal intriguing properties. However, most of them remain for laboratory research, with commercialization still in infancy and yet to be developed. In the future, closer collaboration between academia and industry is anticipated, with increased focus on improving fabrication techniques and product quality, to facilitate more rapid progress toward industrial manufacturing and commercialization of related products.

## III. FABRICATION TECHENIQUES

Over the past two decades, alongside the surge in activities related to 2D materials, many related fabrication techniques have been developed, demonstrating great potential for large-scale industrial manufacturing. In this section, we present an overview of cutting-edge techniques for fabricating integrated photonic devices incorporating 2D materials, highlighting their potential and limitations in industrial manufacturing. It is divided into four parts, including large-scale integration, precise patterning, dynamic tuning, and device packaging.

### A. Large-scale integration

Producing high-quality integrated wafers, such as the silicon-on-insulator (SOI) wafers, is the initial step in manufacturing integrated photonic devices. Current fabrication techniques, such as chemical vapor deposition (CVD) and physical vapor deposition (PVD), have already enabled industrial-scale manufacturing of high-quality integrated wafers in different sizes (*e.g.*, 4, 8, and 12 inches).[40-42] In this part, we focus on discussing methods for integration of 2D materials onto these wafers. To achieve this, the process typically involves two steps: synthesizing 2D materials and transferring them onto integrated substrates. In some cases, 2D materials can be directly synthesized on integrated wafers, thus simplifying the overall process.

Synthesizing high-quality 2D materials is essential for implementing hybrid devices with high performance. Mechanical exfoliation, liquid-phase exfoliation (LPE), and chemical vapor deposition (CVD) are three main methods for synthesizing 2D materials. Mechanical exfoliation provides a simple and efficient way to synthesize 2D materials without using complex facilities. However, for large-scale industrial fabrication, it suffers from limitations due to the inconsistent thicknesses and small sizes of the 2D flakes produced by this method. LPE is a solution-based method to synthesize 2D materials, allowing for cost-effective production of large volumes of 2D nanoflakes. Similar to mechanical exfoliation, it also faces limitations induced by small sizes of the exfoliated 2D flakes. In contrast to the above exfoliation methods, CVD offers an attractive approach for growing high-quality 2D films over large areas with precise thickness control. This makes it appealing for industrial production of high-quality 2D films. Despite this, the state-of-the-art CVD method for synthesizing 2D materials still faces challenges such as limited efficiency, film contamination, and complex process control.

In addition to the above three main methods, some other methods for synthesizing 2D materials have also been investigated, such as molecular beam epitaxy (MBE)[43, 44] and PVD.[45, 46] MBE is a precise atomic layer by layer growth technique used to grow single crystal films epitaxially on an existing crystal substrate. It was invented for the growth of compound semiconductors in the 1970s and has been successfully applied to the growth of metals, oxides, and topological materials.[47] Recently, this technique has also been employed to grow high-quality monolayer graphene, TMDCs, and h-BN.[48-50] The main limitation of current MBE technique is that it requires significantly higher vacuum levels than other deposition methods to achieve comparable impurity levels in the grown films.[51, 52] The PVD method involves a process where material transitions from a condensed phase to a vapor phase, and then recondenses into a thin film. Sputtering and evaporation are two widely used PVD techniques for coating thin films of metal, glass, and polymers in industrial manufacturing. Recently, the PVD method has been employed for fabricating thin hBN, MXenes, and perovskite films. Compared to CVD and MBE, it provides higher deposition rates, but this comes with the trade-off of larger 2D film thicknesses (typically > 30 nm).[45, 46, 53]

After synthesizing 2D materials, it is usually needed to transfer them onto the target integrated substrates to construct hybrid devices. Dry transfer methods using transfer stamps[54] was widely employed for transferring mechanically exfoliated 2D flakes in the early research of this field.[12, 55-57] However, several factors restrict the applicability of this method to laboratory environments instead of industrial production. These mainly include challenges in processing flakes with large lateral size, low fabrication yield, and reliance on complicated supporting facilities. For 2D flakes synthesized by LPE, solution dropping techniques such as drop casting, spin and spray coating are commonly used for their on-chip transfer.[58, 59] Although these methods provide a fast way to prepare 2D films over large areas, they face limitations such as low film uniformity (typically > 10 nm[60]) and large film thicknesses (typically > 100 nm[61]), restricting its use for fabricating films with low thicknesses.

For 2D films synthesized by the CVD method, a high growth temperature is usually needed. As a result, they are typically deposited on metal substrates or foils, rather than dielectric substrates used for integrated photonic devices (*e.g.*, silicon, silicon nitride, silica). Consequently, subsequent film transfer processes are required. Unlike the dry transfer techniques, which are commonly used for small exfoliated 2D flakes, wet transfer techniques are typically employed for transferring CVD-grown 2D films with large lateral sizes.[62, 63] They are easier to operate and do not require sophisticated facilities, resulting in higher success rates and improved transfer efficiency than the dry transfer techniques. Despite these advantages, problems like stretching, wrinkling, and bending can easily occur during the on-chip transfer process, even though CVD-grown 2D films can achieve a high uniformity on the original substrates. This makes it challenging to achieve highly uniform coating or conformal coating on integrated waveguides and metasurfaces, potentially impairing device performance in some applications.[64]

In recent years, several representative works have been reported to address the limitations of the previously mentioned transfer techniques, as summarized in **Fig. 4**. In **Fig. 4(a)**, success transfer of CVD-grown graphene films onto 4-inch silica/silicon wafers was achieved by introducing a transfer medium polymer (polyvinyl alcohol mixed with sorbitol molecules),[65] where the adhesion between the polymer and graphene layers was adjusted through freezing-induced crosslinking, enabling crack-free film transfer over large areas. In **Fig. 4(b)**, a modified solution dropping method was demonstrated by using electrochemical intercalation to prepare high-quality molybdenum disulfide ($MoS_2$) solutions,[66] which allowed the coating of large-area $MoS_2$ films with low thicknesses (~3.6 nm) and high uniformity. In **Fig. 4(c)**, a new solution-based film coating method relying on self-assembly of GO monolayers was demonstrated.[67] This method enabled not only layer-by-layer film coating over large areas on dielectric substrates like silicon, silicon nitride, and silica,[68, 69] but also conformal coatings on integrated waveguides and metasurfaces.[28, 70]

In **Fig. 4(d)**, a tri-layer transfer medium (polymethyl methacrylate (PMMA)/Borneol/graphene) with gradient surface energy was employed to improve the wet transfer process, allowing reliable adhesion to and release from the target substrates.[71] In **Fig. 4(e)**, to maintain 2D films with flat, intact and clean surfaces during the transfer process, supporting films were introduced by incorporating oxhydryl groups-containing volatile molecules into PMMA.[72] These films could deform under heat, thus enabling controllable conformal contact with different 2D films such as graphene and $MoS_2$. In **Fig. 4(f)**, a modified CVD method, combined with a bubble transfer method, was developed to fabricate a 30 × 30 mm single crystal hBN film on a silica/silicon wafer.[73] During the fabrication process, the hBN was delaminated from liquid Au substrates by using hydrogen bubbles generated by electrolysis. This enables the fabrication of high-quality hBN films free from grain boundaries that significantly influence the film uniformity and mechanical strength.

By combining the precise positioning of dry transfer techniques and the high efficiency of wet transfer techniques, a hybrid technique called semi-dry transfer has been developed. For instance, in **Fig. 4(g),** three large graphene sheets of up to 50 × 50 $mm^2$ were transferred to a silicon wafer with over 99% yield.[74] This was achieved by using functional tapes with their adhesive forces controlled by ultraviolet light, along with electrochemical delamination. In **Fig. 4(h)**, the semi-dry transfer method was employed to fabricate a monolayer tungsten disulfide ($WS_2$) on a silicon wafer. During the transfer process, the film was first exfoliated by thermal release tape assisted with a Ni film, followed by releasing the tape/Ni/film on the target wafer. Subsequently, the tape was removed by annealing, and the Ni was etched away using $FeCl_3$.[75] In **Fig. 4(i)**, direct growth of monolayer $MoS_2$ films onto 12-inch silicon wafers was achieved by pre-positioning amorphous $Al_2O_3$ films as interlayers. This provides an effective way for in-situ growing 2D films onto integrated wafers without the film transfer process.[76]

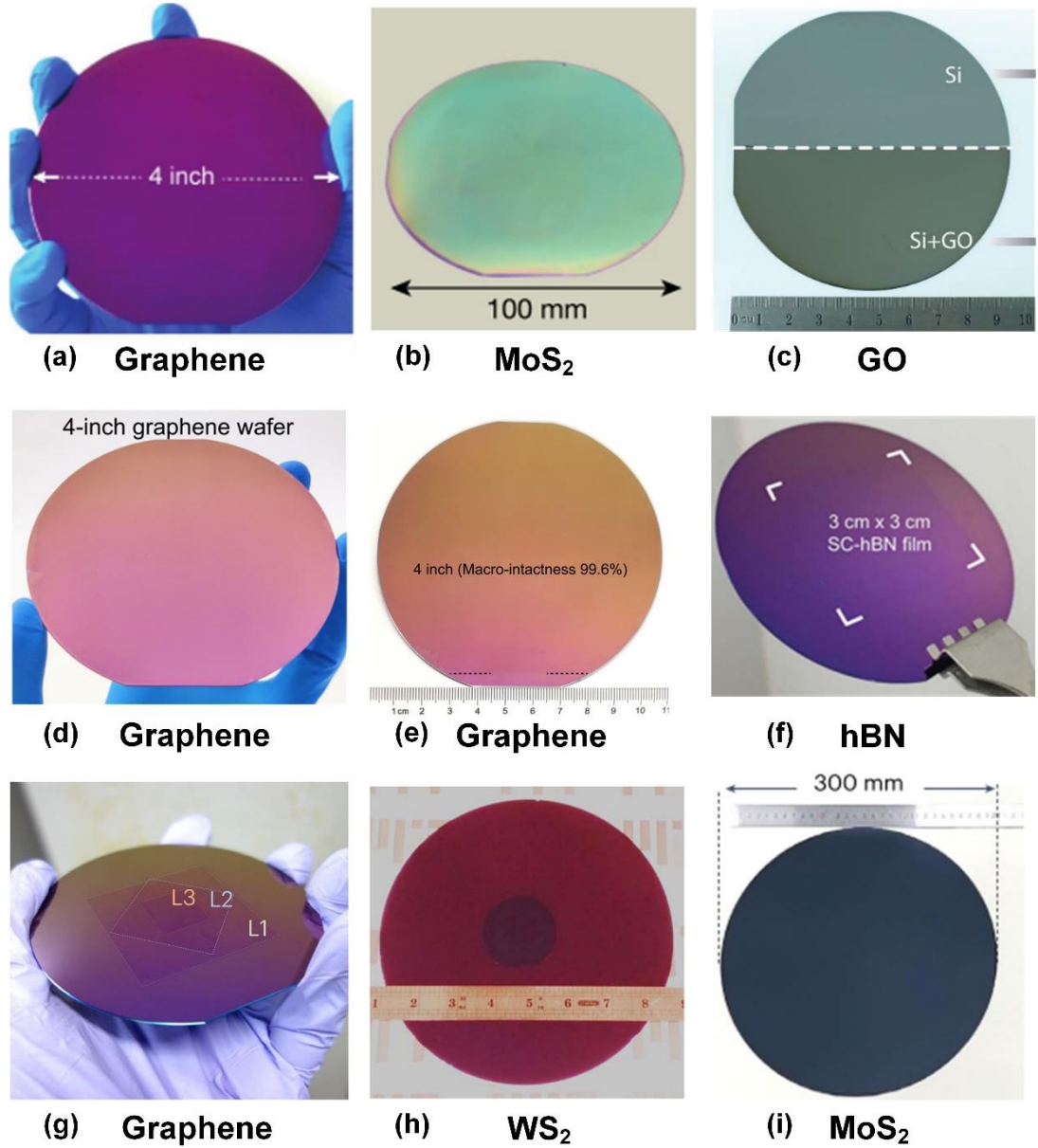

**FIG. 4.** Wafer-scale integrated 2D material films. (a) A graphene film integrated on a 4-inch silicon wafer via dry transfer method. (b) A MoS$_2$ film integrated on a 100-mm diameter silicon wafer via solution dropping method. (c) A GO film integrated on a 4-inch silicon wafer via self-assembly method. (d) A graphene film integrated on a 4-inch silicon wafer via wet transfer method. (e) A graphene film integrated on a 4-inch silicon wafer via bubbling-based wet transfer method. (f) An hBN film integrated on a silicon wafer via wet transfer method. (g) A graphene film integrated on a silica substrate via semi-dry transfer method. (h) A WS$_2$ film integrated on an 8-inch silicon wafer via semi-dry transfer method. (i) A MoS$_2$ film integrated on a 12-inch silicon wafer via direct growth method. (a) Reproduced with permission from Chen *et al.*, Adv. Mater. **36**(15), e2308950 (2024). Copyright 2024 John Wiley and Sons. (b) Reproduced with permission from Lin *et al.*, Nature **562**(7726), 254-258 (2018). Copyright 2018 Springer Nature. (c) Reproduced with permission from Yang *et al.*, ACS Photonics **6**(4), 1033-1040 (2019). Copyright 2019 American Chemical Society. (d) Reproduced with permission from Gao *et al.*, Nat. Commun. **13**(1), 5410 (2022). Copyright 2022 Springer Nature.

(e) Reproduced with permission from Zhao *et al.*, Nat. Commun. **13**(1), 4409 (2022). Copyright 2022 Springer Nature. (f) Reproduced with permission from Lee *et al.*, Science **362**(6416), 817-821 (2018). Copyright 2018 The American Association for the Advancement of Science. (g) Reproduced with permission from Nakatani *et al.*, Nat. Electron. **7**, 119-130 (2024). Copyright 2024 Springer Nature. (h) Reproduced with permission from Shim *et al.*, Science **362**(6415), 665-670 (2018). Copyright 2018 The American Association for the Advancement of Science. (i) Reproduced with permission from Xia *et al.*, Nat. Mater. **22**(11), 5410 (2023). Copyright 2023 Springer Nature.

**Tabel I** summarizes several synthesis and transfer techniques with high potential for large-scale manufacturing (some of which were discussed in **Fig. 4**), along with the relevant publications that utilize these methods. Although the current methods represent significant progress toward industrial manufacturing, challenges still remain. The modified dry transfer method in **Fig. 4(a)** has difficulties in achieving high fabrication yield and accurate film thickness control. The modified solution dropping method in **Fig. 4(b)** has limitations related to reproducibility and solution contamination. The self-assembly method in **Fig. 4(c)** face challenges for fabricating thick (>200 nm) films due to its time-consuming nature, as well as increased scattering loss in thicker films. The performance of the modified wet transfer techniques in **Figs. 4(d) − (f)** can be compromised by challenges such as residual solvents induced by the chemical solution, precise alignment during the transfer, and variability in thickness control. The semi-dry transfer approaches in **Figs. 4(g)** and **4(h)** are currently limited by the complexity of the transfer procedures and potential contamination from chemical solutions. The introduction of $Al_2O_3$ interlayers for the direct growth method in **Fig. 4(i)** was used for fabricating TMDCs, and its applicability to other 2D materials remains to be investigated. In the future, fabrication techniques for large-scale integration of 2D materials can be further developed on the basis of these methods, along with the introduction of new techniques, to ultimately achieve the goal of industrial manufacturing.

**TABLE I.** Synthesis and transfer techniques with high potential for on-chip integration of 2D materials in industrial manufacturing. CVD: chemical vapor deposition, LPE: liquid-phase exfoliation.

| 2D Materials | 2D film thickness | Integrated substrate | Synthesis method | Transfer method | Ref. |
|---|---|---|---|---|---|
| Graphene | Monolayer | 4-inch silicon wafer | CVD | Dry transfer | 65 |
| Graphene | Monolayer | 4-inch silicon wafer | CVD | Wet transfer | 71 |
| Graphene | ~1.6 nm | 4-inch silicon wafer | CVD | Wet transfer | 72 |
| Graphene | Monolayer | silicon wafer | CVD | Semi-dry transfer | 74 |
| GO | ~ 2 μm | SU-8 polymer | LPE | Drop casting | 77 |
| GO | ~1 nm | 4-inch silicon wafer | LPE | Self-assembly | 67 |
| $MoS_2$ | ~0.8 nm | 6-inch silicon wafer | CVD | Wet transfer | 78 |
| $MoS_2$ | ~3.8 nm | 100-mm diameter silicon wafer | LPE | Spin coating | 66 |
| $MoS_2$ | 10.5 − 11.4 nm | 2-inch silicon wafer | LPE | Spin coating | 79 |
| $MoS_2$ | Monolayer | 12-inch silicon wafer | CVD | Direct growth | 76 |
| $WS_2$ | ~ 0.7 nm | 8-inch silicon wafer | CVD | Semi-dry | 75 |
| hBN | Monolayer | silicon wafer | CVD | Wet transfer | 73 |
| MXene | 5–30 layers | silica / silicon nitride | LPE | Self-assembly | 17 |
| MXene | Monolayer | 4-inch silicon wafer | LPE | Spin coating | 80 |

## B. Precise patterning

Precise device patterning is crucial for engineering the functionalities of integrated photonic devices and optimizing their performance. In industrial manufacture of bulk integrated photonic devices, the patterning of dielectric patterns (*e.g.*, waveguides) is mainly performed using photolithography, followed by etching processes such as inductively coupled plasma etching and reactive ion etching. On the other hand, the patterning of metal (*e.g.*, for electrodes and plasmonic devices) is primarily achieved through photolithography, followed by electron beam evaporation and the lift-off processes. Given that the patterning techniques for fabricating bulk integrated photonic devices has been relatively mature,[9, 81-83] here we focus on discussing the methods for patterning 2D materials in hybrid integrated devices. These include not only conventional techniques employed for bulk integrated photonic devices, but also new methods specifically designed for 2D materials.

Photolithography is a dominant device patterning technique in the IC industry. In the photolithography process, the designed layout is transferred from a mask onto photoresist coated on an integrated substrate. Depending on the employed light wavelength, it can be categorized into ultraviolet (UV), deep ultraviolet (DUV), and extreme ultraviolet (EUV) lithography, which allow different levels of patterning resolution.[84, 85] Unlike electron-beam lithography (EBL), which achieves a higher resolution at the expense of considerably longer exposure times, photolithography is more widely employed for industrial manufacturing. In **Fig. 5(a)**, 50-nm-wide graphene nanoribbons featuring straight, clean, and parallel edges were fabricated by using a multi-patterning process that involved photolithography and bottom-up self-expansion. This method enabled a patterning resolution < 100 nm, which is one order of magnitude lower than the traditional lithographical resolution.[86] In **Fig. 5(b)**, a 50-µm-long GO film was patterned on a doped silica micro-ring resonator (MRR). The film placement and coating length were precisely controlled via photolithography on a photoresist, followed by GO film coating (through self-assembly) and lift-off processes.[87]

Nanoimprinting is another widely used patterning technique in integrated device fabrication. Compared to photolithography, it has similarities in steps like resist coating, film attachment, and resist removal, but differs by using an imprint mold to pattern the resist layer instead of photolithography. In **Fig. 5(c)**, graphene nanoring arrays were fabricated on a 9-cm$^2$ silica substrate via nanoimprinting. The fabricated rings had a width of <15 nm and a thickness of <0.7 nm.[88] In addition, nanoimprinting was utilized to engineer the strain of 2D MoS$_2$ films in **Fig. 5(d)**, allowing for precise control of the strain magnitudes and distributions at a low cost and high throughput.[89]

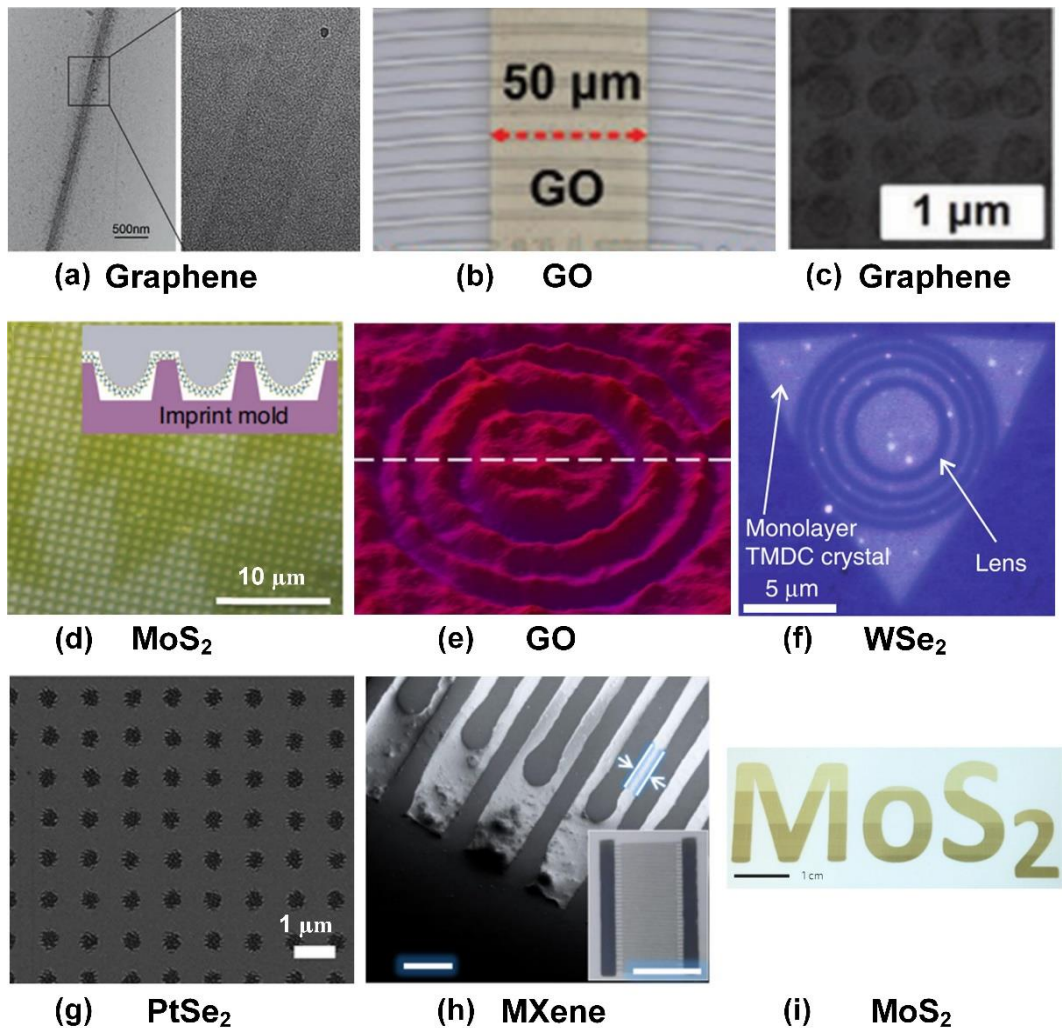

**FIG. 5.** Methods for patterning 2D material films on integrated substrates. (a) A graphene nanoribbon patterned via photolithography and bottom-up self-expansion. (b) A GO film patterned via photolithography and lift-off. (c) Graphene nanoring arrays patterned via nanoimprinting. (d) A $MoS_2$ film patterned via nanoimprinting. (e) A 200-nm-thick GO lens patterned via laser patterning. (f) A monolayer tungsten diselenide ($WSe_2$) lens patterned via laser patterning. (g) $PtSe_2$ nanohole arrays patterned via laser patterning. (h) MXene lines patterned via inkjet printing. (i) '$MoS_2$' word printed via inkjet printing. (a) Reproduced with permission from Borhade *et al.*, Small **20**(22), 2311209 (2024). Copyright 2024 John Wiley and Sons. (b) Reproduced with permission from Wu *et al.*, Small **16**(16), 1906563 (2020). Copyright 2020 John Wiley and Sons. (c) Reproduced with permission from Pak *et al.*, Adv. Mater. **25**(2), 199-204 (2013). Copyright 2013 John Wiley and Sons. (d) Reproduced with permission from Sun *et al.*, Microsyst. Nanoeng. **10**(49) (2024). Copyright 2024 Springer Nature. (e) Reproduced with permission from Zheng *et al.*, Nat. Commun. **6**, 8433 (2015). Copyright 2015 Springer Nature. (f) Reproduced with permission from Lin *et al.*, Light Sci. Appl. **9**, 137 (2020). Copyright 2020 Springer Nature. (g) Reproduced with permission from Enrico *et al.*, ACS Nano **17**(9), 8041-8052 (2023). Copyright 2023 American Chemical Society. (h) Reproduced with permission from Zhang *et al.*, Nat. Commun. **10**(1), 1795 (2019). Copyright 2019 Springer Nature. (i) Reproduced with permission from McManus *et al.*, Nat. Nanotechnol. **12**(4), 343-350 (2017). Copyright 2017 Springer Nature.

Compared to photolithography and nanoimprinting, laser patterning provides a quick and simple method for directly patterning 2D film without using any masks, molds, photoresists, or chemical solutions. In **Fig. 5(e)**, a GO flat lens was fabricated by using laser patterning to convert the GO into reduced GO (rGO) via photoreduction, achieving a focusing resolution of ~300 nm and an absolute focusing efficiency >32% over a broad wavelength range from 400 to 1500 nm.[90] In **Fig. 5(f)**, laser patterning was used for fabricating a flat lens on monolayer TMDC, achieving a subwavelength lateral resolution of ~400 nm and a high focusing efficiency of ~31%.[91] In **Fig. 5(g)**, a noncontact laser patterning method was proposed by using a two-photon 3D printer with a scanning pulsed laser. 2D materials like graphene, $MoS_2$, and platinum diselenide ($PtSe_2$) were patterned with a subwavelength resolution (*e.g.*, ≥100-nm hole diameter) at a high throughput (*e.g.*, ~3s to clear a 200 μm × 200 μm area).[92]

As a modified solution dropping technique, inkjet printing enables the on-chip transfer and patterning of 2D films in a single step. Its fast and in-situ patterning allows for industrial fabrication of large-area patterns, with their position and shape controlled by program. In **Fig. 5(h)**, additive-free MXene ink for was used for inkjet printing, achieving a line width of ~80 μm, a gap width of ~50 μm, and a spatial uniformity of ~3.3%.[93] In **Fig. 5(i)**, inkjet printing was used for patterning $MoS_2$ and graphene in large areas, with an optimal drop spacing of ~40 μm on silicon/silica substrates and ~45 μm on glass and polyethylene terephthalate.[94]

Although the aforementioned methods have been extensively used for patterning 2D materials in laboratory research and offer appealing advantages for industrial fabrication, challenges still remain. For example, photolithography has limitations induced by chemical residues after the etching or lift-off process, damage in 2D films due to high-energy exposure or aggressive chemical solutions, and reduced effectiveness on non-planar substrates. The laser patterning process can also cause damage to 2D films. In addition, the minimum feature size of laser patterning is limited by the spot size of the focused laser beam, resulting in a relatively low patterning resolution that is typically > 300 nm. Nanoimprinting face challenges related to its flexibility in changing the device pattern and in achieving a high film uniformity across large areas. These make it better suited for fabricating relatively simple and repetitive patterns. The imprinting process can also cause mechanical damage to the delicate structures of 2D materials, thus degrading their performance for some applications. For inkjet printing, the patterning resolution is relatively low (typically > 1 μm), and it also face challenges similar to other solution dropping methods, such as low film uniformity and large film thicknesses.

Some other patterning methods, such as focused ion beam (FIB) milling,[95-97] scanning probe lithography (SPL),[98, 99] adhesion lithography,[100] and self-assembled-mask lithography (SAML),[101, 102] have also been employed for patterning 2D materials. Each method has its own distinct advantages, such as the high patterning resolution for FIB milling (*e.g.*, down to 10 nm[95]) and the rapid processing speed for SPL (*e.g.*, ~20 mm/s[103]). However, either relatively low fabrication efficiency or limited applicability to different 2D materials restrict their current use to laboratory research rather than industrial-scale production. With continued progress in micro/nano fabrication techniques, it is anticipated that more advanced patterning methods aimed at low-cost and highly efficient industrial manufacturing would be developed to overcome existing limitations in the future.

## C. Dynamic tuning

Dynamic tuning underpins the operation of many active integrated photonic devices such as lasers,[10, 104] optical switches,[105, 106] and electro-optic modulators.[107, 108] Even for passive integrated photonic devices, introducing dynamic tuning is useful for optimizing their performance[87, 109] and extending their applicability across varying conditions.[110, 111] For commercial bulk integrated photonic devices, dynamic tuning is typically achieved by introducing co-integrated thermo-optic micro-heaters[112, 113] or PN junctions[114, 115] to change material refractive indices. The former has typical response times in the millisecond to microsecond range, whereas the latter enables faster tuning with response times on the order of picoseconds or even faster. In laboratory research, dynamic tuning can also be achieved through laser-induced photothermal effects or nonlinear optical effects.[116, 117] For hybrid integrated photonic devices incorporating 2D materials, dynamic tuning of the properties of 2D materials is also crucial for enabling functionalities and optimizing performance. In this part, we discuss the mechanisms for dynamic tuning of 2D materials in hybrid integrated photonic devices, which are classified into gate tuning, laser tuning, thermal tuning, and strain tuning.

2D materials exhibiting metallic behaviors, such as graphene and TMDCs, are susceptible to external electrical fields due to their exceptional Fermi-Dirac tunability. This forms the basis of electrostatic gate tuning methods, which allows for efficient, reversible, and real-time modulation of their electrical and optical properties. In **Fig. 6(a)**, a graphene/ion-gel heterostructure was integrated on a silicon nitride MRR with source–drain and top gating, which enabled gate tuning of the Fermi level of graphene and hence the chromatic dispersion of the hybrid MRR.[118] In **Fig. 6(b)**, a monolayer $WS_2$ was patterned on a silicon nitride MRR with ionic liquid cladding. The $WS_2$ film was doped by applying a bias voltage across the two electrodes through the ionic liquid, resulting in a significant refractive index change of $\Delta n$ = 0.53.[119] In **Fig. 6(c)**, ferroelectric gate tuning was achieved by co-integrating a ferroelectric P(VDF-TrFE) top layer over a 2D molybdenum ditelluride ($MoTe_2$) film on a silicon/silica substrate. By using a scanning probe to control the polarization of the ferroelectric polymers, lateral p–n, n–p, p–p, n–n homojunctions were constructed in the $MoTe_2$ film, allowing simple and arbitrary definition of the carrier injection with nanoscale precision.[120]

Lasers can be used to trigger optical effects like saturable absorption, photon-excited carrier transport, and photothermal effects in 2D materials, which can, in turn, be harnessed to dynamically tune their properties. In **Fig. 6(d)**, a graphene was integrated on a silicon photonic crystal cavity, where the optical absorption of graphene was tuned via exposure to a continuous-wave (CW) laser at 1064 nm, achieving a resonance wavelength shift of ~3.5 nm and a quality factor change of ~20%.[121] In **Fig. 6(e)**, the light absorption in a graphene-silicon hybrid nanowire waveguide was tuned by exposure to a 635-nm CW laser. During the tuning process, the photon-excited carriers in silicon were injected into graphene via the Schottky diode junction, resulting in an increased carrier concentration and change of the Fermi level.[122] In **Fig. 6(f)**, by injecting a CW light into GO-silicon hybrid waveguides to induce reversible photo-thermal changes in 2D GO films, three functionalities were demonstrated across broad wavelength ranges, including all-optical control and tuning, optical power limiting, and non-reciprocal light transmission.[123]

Apart from laser-induced photo-thermal changes, co-integrated micro-heaters can also induce photothermal changes in 2D materials. Unlike the hybrid tuning approach in **Fig. 6(f)** that combines laser and thermal tuning, the use of micro-heaters to cause temperature-dependent changes is a typical thermal tuning method. In **Fig. 6(g)**, a ring-shaped metallic microheater was fabricated on top of polypropylene, with a $MoS_2$ film placed at its center. By applying a bias voltage to the microheater, the temperature of the polypropylene surface was changed, enabling precise and reversible tuning of the refractive index of $MoS_2$.[124]

Strain tuning provides an effective method to modify the properties of 2D materials in a continuous and reversible manner by applying mechanical tension or compression. Localized strain can be utilized to finely control and adjust material properties such as optical emission and photoconductivity in 2D materials. However, in comparison to gate or optical tuning, the response time of strain tuning is relatively slow. In **Fig. 6(h)**, single photon emission in a 2D hBN film were controlled via local strain tuning induced by atomic force microscope indentation. By tuning the indentation parameters, indents sites with various lateral sizes were obtained to fully activate the single-photon emitters.[125] In **Fig. 6(i)**, single-photon emitters were achieved in a $WSe_2$ film by using strain tuning through nanoscale stressors combined with defect engineering via electron-beam irradiation. This method not only allowed precise tuning of the emission sites but also improved the yield, purity, and operational temperature of the emitters.[126]

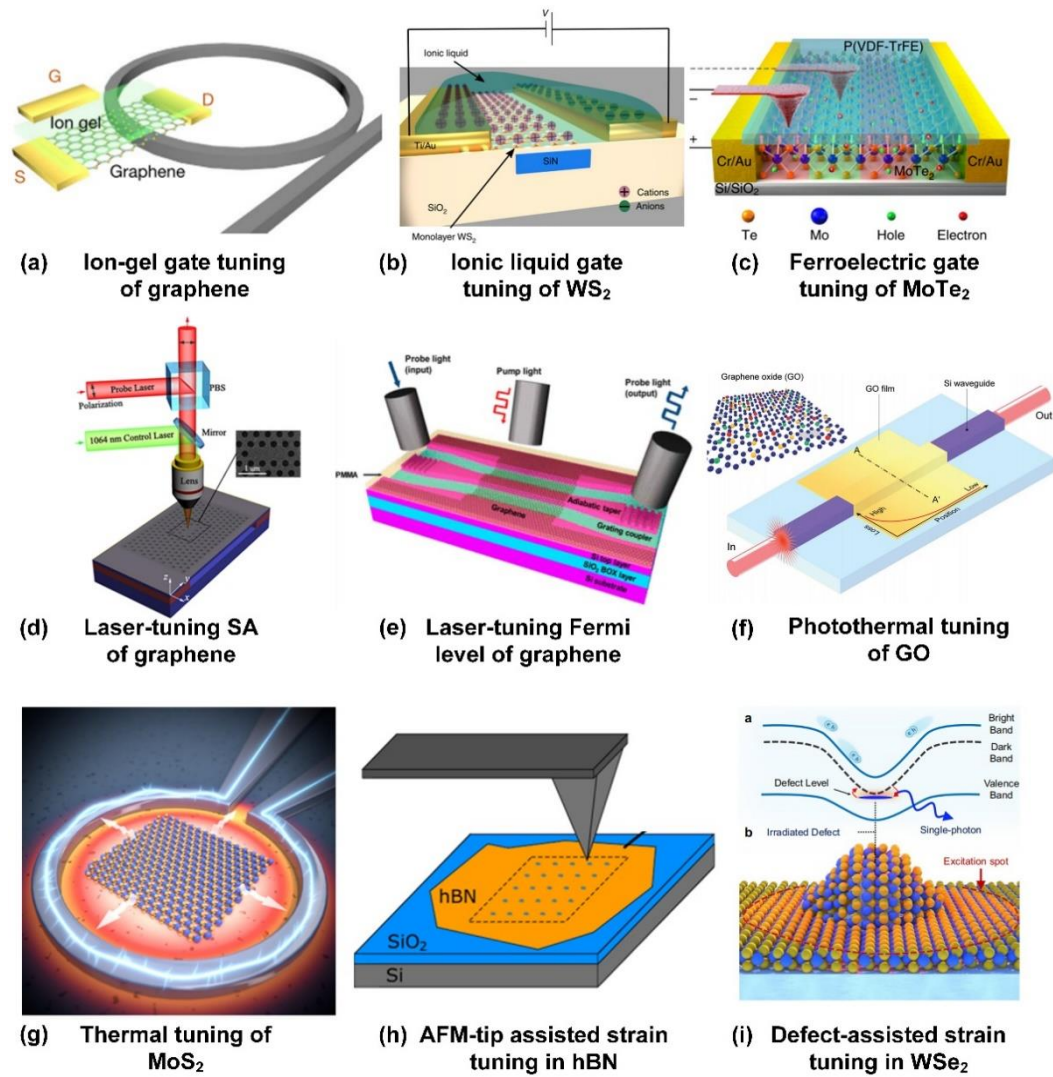

**FIG. 6.** Methods for dynamic tuning of 2D materials in integrated photonic devices. (a) Ion-gel gate tuning of graphene integrated on a silicon nitride MRR. (b) Ionic liquid gate tuning of $WS_2$ integrated on a silicon nitride MRR. (c) Ferroelectric gate tuning of $MoTe_2$ integrated on a silicon substrate. (d) Laser tuning saturable absorption (SA) of graphene integrated on a silicon photonic crystal cavity. (e) Laser tuning Fermi level of graphene integrated on silicon-on-insulator nanowires. (f) Photothermal tuning of GO integrated on a silicon waveguide. (g) Thermal tuning of $MoS_2$ integrated on a polypropylene substrate. (h) Atomic force microscopy (AFM)-assisted strain tuning of hBN integrated on a silicon substrate. (i) Defect-assisted strain tuning of $WSe_2$ integrated on a silicon substrate. (a) Reproduced with permission from Yao *et al.*, Nature **558**(7710), 410-414 (2018). Copyright 2018 Springer Nature. (b) Reproduced with permission from Datta *et al.*, Nat. Photonic **14**(4), 256-262 (2020). Copyright 2020 Springer Nature. (c) Reproduced with permission from Wu *et al.*, Nat. Electron. **3**, 43-50 (2020). Copyright 2020 Springer Nature. (d) Reproduced with permission from Shi *et al.*, ACS Photonics **2**(11), 1513-1518 (2015). Copyright 2015 American Chemical Society. (e) Reproduced with permission from Yu *et al.*, ACS Photonics **8**(11), 11386–11393 (2014). Copyright 2014 American Chemical Society. (f) Reproduced with permission from Wu *et al.*, Adv. Mater. **36**(16), 2403659 (2024). Copyright 2024 John Wiley and Sons. (g) Reproduced with permission from Ryu *et al.*, Nano Lett. **20**(7), 5339-5345 (2020). Copyright 2020

American Chemical Society. (h) Reproduced with permission from Xu *et al*., Nano Lett. **21**(19), 8182-8189 (2021). Copyright 2021 American Chemical Society. (i) Reproduced with permission from Parto *et al*., Nat. Commun. **12**(1), 3585 (2021). Copyright 2021 Springer Nature.

Although the above dynamic tuning methods of 2D materials offer attractive benefits, several challenges still need to be addressed for their use in industrial fabrication. For instance, traditional gating tuning methods often struggle to achieve precise modulation in small, localized regions. This lack of precision can limit the effectiveness of tuning, especially for nanoscale devices. For laser tuning method, the tuning lasers are usually not co-integrated on chips, which hinders its wide use for commercial products. In addition, improper laser intensity or wavelength can result in localized overheating or photodegradation of 2D materials, leading to irreversible damage such as ablation or defect formation. Some strain tuning methods are prone to damage the 2D materials during the tuning process such as cracking, folding, or tearing, especially when the strain exceeds the material elastic limit. The relatively low response speed of strain tuning also limits its use in high-speed tuning applications. To address these challenges, it is expected that future advancements in dynamic tuning methods would be developed with more robust, flexible, and versatile control.

## D. Packaging

Device packaging typically represents the final stage of integrated device fabrication, where the device is encapsulated in a protective case that safeguards it from physical damage and corrosion. This step is essential for the industrial manufacturing of commercial products with stable operation. After over 50 years of development, the IC industry has already established well-developed packaging techniques, such as pin grid array (PGA), land grid array (LGA), ball grid array (BGA) and 2D & 3D customized solutions.[127-130] With respect to PICs, some advanced packaging techniques have been adapted from those used for ICs, while also taking into account new factors such as fiber-to-chip coupling, chip-to-chip coupling, high-density electrical and optical interconnections, hybrid integration of photonic chips, multi-chip modules, and thermal management. **Figs. 7(a) – (c)** show some images that showcase state-of-the-art packaging techniques for integrated photonic devices. **Fig. 7(a)** shows a hybrid tunable laser butterfly packaged by the LioniX International corporation, which includes active gain medium and a passive external cavity.[131] **Fig. 7(b)** shows a packaging technique developed by PHIX corporation, which utilizes off-the-shelf building blocks and provides an open architecture that accommodates devices with various sizes and across different integrated platforms.[132] **Fig. 7(c)** shows the first-ever fully integrated optical compute interconnect chiplet co-packaged with a Central Processing Unit (CPU) from the Intel corporation.[133]

For hybrid integrated photonics devices incorporating 2D materials, in addition to packaging integrated chips, encapsulating 2D materials is also necessary to prevent material degradation and ensure stable operation. Due to their large surface area and ultralow film thicknesses, many 2D materials (*e.g.*, BP and TMDCs) are highly sensitive to environmental factors such as temperature, humidity, mechanical stress, and chemical exposure. To improve their stability in hybrid integrated devices, various encapsulation materials, such as dielectric materials, organic polymers, and robust 2D materials, have been used to effectively shield 2D materials from the surrounding environment.

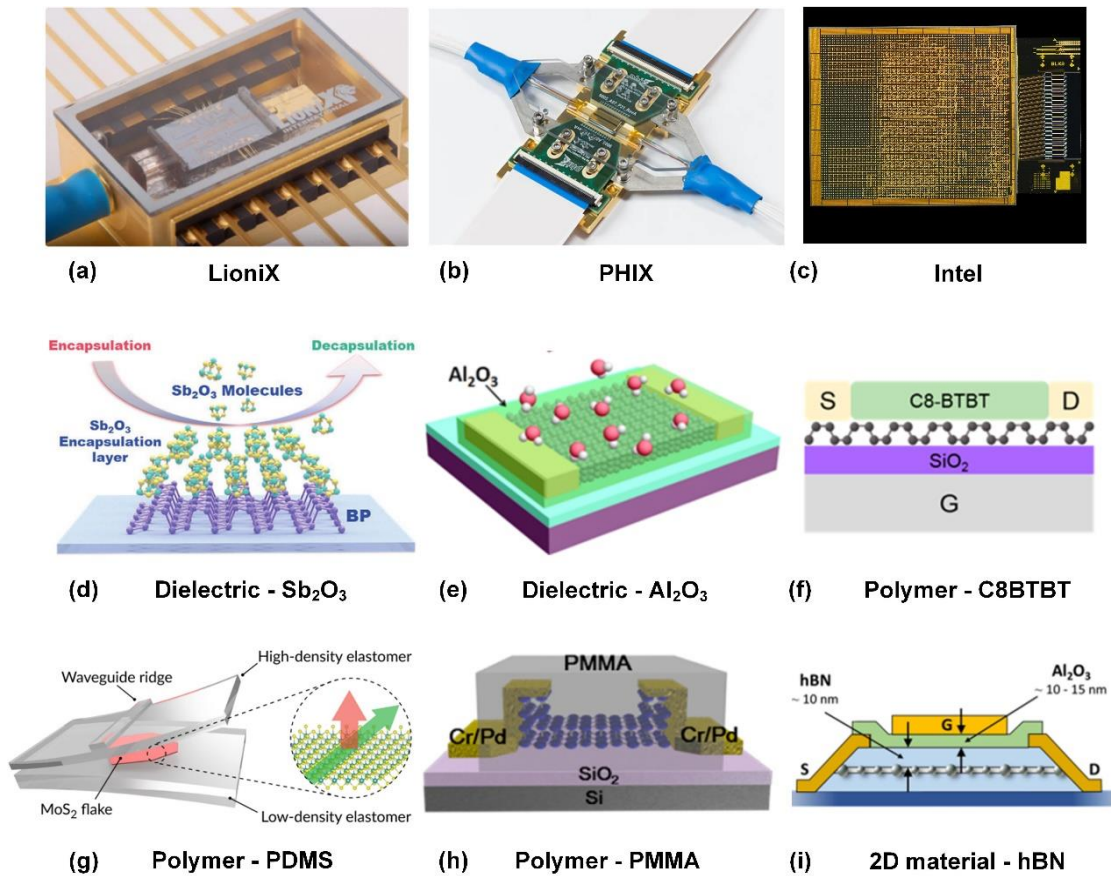

**FIG. 7.** Methods for packaging PICs and encapsulation of 2D materials in hybrid integrated photonic devices. (a) A hybrid tunable laser packaged by the LioniX International corporation. (b) A packaging module for photonic devices with an open architecture from PHIX corporation. (c) An integrated optical compute interconnect chiplet packaged by the Intel corporation. (d) Encapsulation of a BP film by covering a 3-nm-thick inorganic molecular crystal $Sb_2O_3$ layer. (e) Encapsulation of a BP film by covering a 6-nm-thick $Al_2O_3$ layer. (f) Encapsulation of a BP film by covering a 60-nm-thick C8-BTBT organic film. (g) Encapsulation of a $MoS_2$ film in a polydimethylsiloxane (PDMS) waveguide chip. (h) Encapsulation of a BP film by covering a 300-nm-thick polymethyl methacrylate (PMMA) polymer layer. (i) Encapsulation of a graphene film between two hBN layers. (a) Reproduced from https://www.lionix-international.com/photonics/pic-technology/assembly-and-packaging-service/. (b) Reproduced from https://www.phix.com/our-offering/prototype-package/phix-characterization-package/. (c) Reproduced from https://www.intel.com/content/www/us/en/newsroom/news/intel-unveils-first-integrated-optical-io-chiplet.html#gs.gpexcr. (d) Reproduced with permission from Liu *et al.*, Adv. Mater. **34**(7), e2106041 (2022). Copyright 2022 John Wiley and Sons. (e) Reproduced with permission from Miao *et al.*, ACS Appl. Mater. Interfaces **9**(11), 10019-10026 (2017). Copyright 2017 John Wiley and Sons. American Chemical Society. (f) Reproduced with permission from He *et al.*, Nano Lett. **19**(1), 331-337 (2019). Copyright 2019 American Chemical Society. (g) Reproduced with permission from Auksztol *et al.*, ACS Photonics **6**(3), 595-599 (2019). Copyright 2019 American Chemical Society. (h) Reproduced with permission from Jia *et al.*, ACS Nano **9**(9), 8729-8736 (2015). Copyright 2015 American Chemical Society. (i) Reproduced with permission from Viti *et al.*, Nano Lett. **20**(5), 3169-3177 (2020). Copyright 2020 American Chemical Society.

In **Fig. 7(d)**, by using 3-nm-thick inorganic molecular crystal $Sb_2O_3$ as a protective layer, BP exhibited significantly enhanced structural stability for over 80 days, in contrast to the rapid degradation of unprotected BP within hours.[134] In **Fig. 7(e)**, a 6-nm-thick $Al_2O_3$ encapsulation layer was deposited on a BP film via atomic layer deposition, enabling the realization of air-stable BP-based humidity /chemical sensors.[135] In **Fig. 7(f)**, the stability of BP was improved by using a 60-nm-thick dioctylbenzothienobenzothiophene (C8-BTBT) organic film, which provided protection against oxidation under ambient conditions for more than 20 days. This approach can also be applied to other 2D materials such as TMDCs.[136] In **Fig. 7(g)**, a monolayer of $MoS_2$ was embedded within an elastomeric waveguide chip made from polydimethylsiloxane (PDMS), which not only ensured mechanical and environmental protection of the $MoS_2$ film but also enhanced its photoluminescence (PL) performance.[137] In **Fig. 7(h)**, BP films were plasma-treated and passivated with a ~300-nm-thick PMMA polymer cover layer, which not only eliminated chemical degradation of the oxidized BP surface but also provided protection against water and oxygen molecules in the air.[138] In **Fig. 7(i)**, a monolayer graphene film was encapsulated between two hBN layers (~30 nm on the bottom and ~10 nm on the top) with high thermal, chemical, and mechanical stability, allowing for broadband and low-noise terahertz photodetection at room temperature.[139]

Although the above encapsulation methods provide many advantages in protecting 2D materials from environmental degradation, they still present some limitations in terms of broad applicability and long-term performance needed for commercial products. For instance, physical encapsulation with polymers like PMMA suffers from limited durability and will degrade rapidly when exposed to organic solvents. In addition, the encapsulation with dielectric materials such as $Al_2O_3$ via atomic layer deposition method and chemical functionalization with organic solutions is destructive, leading to inevitable degradation of the properties of 2D materials. The encapsulation methods such as dielectrics deposition and polymer coatings may also compromise the sensitivity and functionality of 2D materials in integrated photonic devices, resulting in a trade-off between protection and performance. Therefore, it is expected that more non-invasive encapsulation methods would be developed in the future, not only providing enhanced protection for 2D materials against environmental degradation, but also allowing their potential to be fully exploited.

## IV. COMMERCIALIZATION ISSUES

To ensure efficient manufacturing of reliable commercial products, it's necessary to consider several factors beyond fabrication techniques. In this section, we discuss some issues related to commercialization of integrated photonic devices incorporating 2D materials. It is divided into four parts, including fabrication standards, product recycling, service life, and environmental implications.

### A. Fabrication standards

In industrial manufacturing, establishing clear and detailed fabrication standards is crucial for ensuring product consistency and reproducibility. For bulk integrated photonic devices, comprehensive fabrication standards have already been established, such as IEC Technical Report 63072 by the International Electrotechnical Commission (IEC)[140] and ISO-10110 by the International Organization for Standardization (ISO).[141] However, for integrated photonic devices incorporating 2D materials, there is still a lack of unified protocols or fabrication standards in the manufacturing processes.

Since the properties of 2D materials are affected by the fabrication methods, the quality and consistency of synthesized materials in practical settings vary widely, which has limited its large-scale industrial applications. To enable accurate evaluations and comparisons of the quality of 2D materials, it is essential to establish widely recognized and quantitative certification standards. In this part, we discuss the possibility of establishing fabrication standards for evaluating the quality of 2D materials in hybrid integrated devices. These standards can guide the selection and characterization of 2D materials for industrial manufacturing, ensuring that they meet specific performance criteria necessary for commercial products.

**TABLE II.** Typical characterization methods for assessing quality and properties of 2D materials. XPS: X-ray photoelectron spectroscopy; XRD:X-ray diffraction; UV-VIS: ultraviolet-visible; AFM: atomic force microscopy; SEM: scanning electron microscopy; TEM: transmission electron microscopy; HRTEM: high-resolution transmission electron microscopy; STEM: scanning transmission electron microscopy; PL: photoluminescence.

| Methods | Main features | Ref. |
|---|---|---|
| Raman spectroscopy | Analyzing chemical composition, molecular vibrations, crystal structure, and defect density by measuring the shift in laser photon energy caused by Raman scattering during light-matter interaction | 142 |
| XPS | Analyzing elemental composition and chemical states by measuring the energy of photoelectrons emitted from the surface after X-ray irradiation | 143 |
| XRD | Analyzing crystal structure and phase by measuring the angles of X-ray diffraction and the intensities of the resulting diffraction peaks | 144 |
| UV-VIS spectrometry | Measure the absorption of ultraviolet and visible light passing through a sample to identify and quantify various compounds | 145 |
| AFM | Topographic imaging to characterize film thickness / uniformity, and force measurement to characterize sample mechanical properties | 146 |
| Optical microscopy | Imaging sample surface by using a system of optical lenses, the maximum magnification typically ranges from 500X to 1500X | 147 |
| SEM | Imaging sample surface by scanning a focused electron beam, the maximum magnification typically ranges from 10,000X to 500,000X | 148 |
| TEM | Imaging internal structure at atomic level by transmitting a beam of high-energy electrons through thin samples, the maximum magnification typically ranges from 100,000X to 1,000,000X | 149 |
| HRTEM | A high-resolution imaging mode of specialized TEM that allows for direct imaging of the atomic structure of samples | 150 |
| STEM | Combining SEM and TEM for high-resolution atomic-scale imaging | 151 |
| PL | Measuring light emission from materials under optical excitation to analyze optical properties of samples | 152 |
| Carrier mobility | Measuring mobility of charge carriers when subjected to external electric fields to characterize electrical properties of samples | 153 |

**Table II** provides an overview of typical methods for characterizing the quality and properties of 2D materials. In laboratory research, Raman spectroscopy is the most widely used approach for characterizing the properties of 2D materials. By measuring the shift in laser photon energy induced by Raman scattering, it can be used to analyze the molecular vibrations, crystal structure, chemical composition, and defect density in samples. For example, the intensity ratios of $D$ to $G$ peaks ($I_D/I_G$) obtained from the Raman spectra are useful for characterizing the properties of graphene family materials. In **Fig. 8(a)**, the change of $I_D/I_G$ in the measured Raman spectra indicates different levels of reduction in a GO film after applying different input light powers.[154] In addition, the quality of TMDCs can be assessed by analyzing the variations of $E_{2g}$ and $A_{1g}$ peaks in Raman spectra. In **Fig. 8(b)**, after oxygen plasma exposure treatment on pristine monolayer MoS$_2$, the intensities of both $E_{2g}$ and $A_{1g}$ peaks show a significant decrease, indicating a substantial increase in lattice distortion in the MoS$_2$ crystal.[155]

In addition to Raman spectroscopy, X-ray photoelectron spectroscopy (XPS), X-ray diffraction (XRD), and ultraviolet-visible (UV-VIS) spectrometry are also useful methods for characterizing the properties of 2D materials. XPS is a quantitative spectroscopic method that generates electron population spectra by irradiating a material with a beam of X-rays. It can be used to characterize the elemental composition of a material and its chemical state. **Fig. 8(c)** shows an XPS spectra for MoS$_2$ films synthesized by a modified tetraheptylammonium bromide (THAB)-exfoliated method compared with a traditional lithium (Li)-exfoliated method.[66] Analyzing the $3d$ peaks for Mo and $2s$ peak for S reveals that the modified method yields a pure 2H phase of MoS$_2$, whereas the Li-exfoliated MoS$_2$ nanosheets are predominantly in the 1T phase. XRD is widely used to characterize crystal structure and phase by measuring the angles of X-ray diffraction and the intensities of the resulting diffraction peaks. **Fig. 8(d)** shows a real-time peak shift to a higher angle in the XRD spectra for a Ti$_3$C$_2$T$_x$ MXene film, indicating a decreased interlayer spacing due to the removal of water and adsorbents after $N_2$ purging.[156] UV-VIS spectrometry measures the absorption spectra of a sample as ultraviolet and visible light passes through it. This technique is widely employed to identify and quantify various compounds in a variety of samples. Compared to XPS, UV-VIS can analyze the properties of samples in both film and solution forms. **Fig. 8(e)** shows UV-VIS spectra to qualitatively analyze the percentage of few-layer GO sheets in dispersions by

comparing the absorption peak intensities.[157]

Although the above material characterization methods provide useful tools for quantitatively analyze the atomic structure and chemical characteristics of 2D materials, they have limitations in accurately assessing physical characteristics such as film thickness, surface uniformity and adhesion between 2D materials and integrated substrates. To characterize the film thickness and uniformity, atomic force microscopy (AFM) can be employed. **Fig. 8(f)** shows an AFM image of a monolayer GO film fabricated by the solution-based self-assembly method, which had a thickness of ~1 nm and a high uniformity.[67] In addition, some other microscopy methods, such as optical microscopy, scanning electron microscopy (SEM), transmission electron microscopy (TEM), high-resolution transmission electron microscopy (HRTEM), scanning transmission electron microscopy (STEM), and scanning tunneling microscopy (STM), can be used to examine surface morphology, wrinkles, and interface characteristics between different materials. **Fig. 8(g)** shows an SEM image of an SOI nanowire waveguide coated with monolayer of GO, confirming the conformal coating of the 2D films onto the waveguide.[70] **Fig. 8(h)** shows a cross-sectional STEM image of the interface between stacked Au electrodes and 1T'-$ReS_2$ films, clearly indicating a gap of ~3.7 Å at the interface.[158]

Another widely used approach to assessing the quality of 2D materials is to characterize their optical or electrical properties. For instance, the measured PL spectra in **Fig. 8(i)** show a significant quenching in the overlapping area between $MoS_2$ and $WS_2$, which indicates strong interaction between the two materials and the clean interface of the 2D heterostack.[74] In **Fig. 8(j)**, the carrier mobilities of graphene was measured after transferring onto silicon substrates via a modified wet transfer method, ranging between 70,000 and 120,000 cm² V$^{−1}$ s$^{−1}$ at room temperature. These values surpass the standard value of ~30,000 cm² V$^{−1}$ s$^{−1}$ for conventional wet transfer methods, reflecting the high-quality of graphene transferred by this new method.[72]

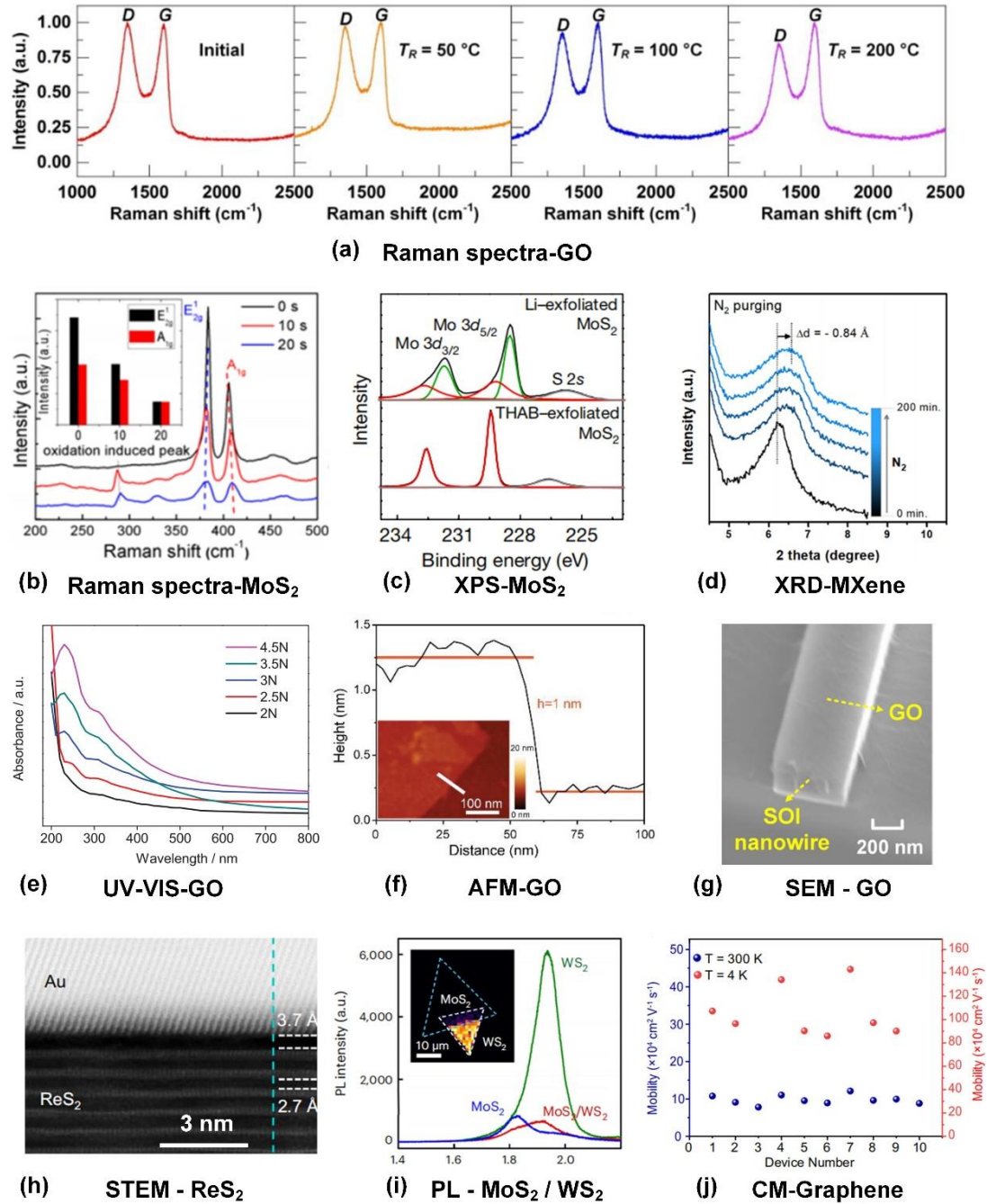

**FIG. 8.** Methods for evaluating quality of 2D materials. (a) Raman spectra of GO at different levels of reduction. (b) Raman spectra of MoS$_2$ before and after oxygen plasma exposure treatment. (c) X-ray photoelectron spectra (XPS) of MoS$_2$ synthesized by two different liquid-phase exfoliation methods. (d) X-ray diffraction (XRD) spectra of Ti$_3$C$_2$T$_x$ MXene film after $N_2$ purging. (e) Ultraviolet-visible (UV-VIS) spectra of GO dispersions. (f) Atomic force microscopy (AFM) image of a monolayer GO film. (g) Scanning electron microscopy (SEM) image of monolayer GO conformally coated on a silicon waveguide. (h) Scanning transmission electron microscopy (STEM) image of the interface between stacked Au electrodes and 1T'-ReS$_2$ films. (i) Photoluminescence (PL) spectra in the overlapping area between MoS$_2$ and WS$_2$. (j) Carrier mobilities (CM) of graphene after being transferred onto silicon substrates. (a) Reproduced with permission from Hu *et al.*, Adv. Funct. Mater. 2406799 (2024). Copyright

2024 John Wiley and Sons. (b) Reproduced with permission from Ye *et al*., Nano Lett. **16**(2), 1097-103 (2016). Copyright 2016 American Chemical Society. (c) Reproduced with permission from Lin *et al*., Nature **562**(7726), 254-258 (2018). Copyright 2018 Springer Nature. (d) Reproduced with permission from Lai *et al*., AIP Advances **2**(3), 032146 (2012). Copyright 2012 AIP Publishing. (e) Reproduced with permission from Yang *et al*., ACS Photonics **6**(4), 1033-1040 (2019). Copyright 2019 American Chemical Society. (f) Reproduced with permission from Zhang *et al*., ACS Appl. Mater. Interfaces **12**(29), 33094-33103 (2020). Copyright 2020 American Chemical Society. (g) Reproduced with permission from Zhang *et al*., Nat. Commun. **15**(1), 4619 (2024). Copyright 2024 Springer Nature. (h) Reproduced with permission from Nakatani *et al*., Nat. Electron. **7**, 119-130 (2024). Copyright 2024 Springer Nature. (i) Reproduced with permission from Zhao *et al*., Nat. Commun. **13**(1), 4409 (2022). Copyright 2022 Springer Nature.

Now the aforementioned methods are mainly used for laboratory research and need to address some limitations before they can be applied more widely in industrial production. For instance, Raman spectroscopy requires high film uniformity when characterizing ultrathin (< 10 nm) 2D films. Increased film roughness can lead to higher scattering loss, which in turn impacts the accuracy of the measurements. In addition, although some advanced instruments like TEM, HRTEM, STEM, and STM can offer exceptional accuracy in characterizing 2D materials, they often come with high costs for acquisition, operation, and maintenance, potentially limiting their widespread adoption in industrial environments. Testing optical and electrical properties of 2D films involve more complex procedures and specialized instruments. It is also sensitive to temperature fluctuations, which can lead to inconsistent results in varying conditions. Characterizing each individual device can be a time-consuming process. Therefore, extensive sampling and statistical analysis are required during large-scale production to guarantee consistent material quality across various batches. In the future, it is expected that some more general and cost-effective methods such as Raman spectroscopy and AFM will initially be employed in industrial production, and these methods will gradually evolve into automatic and high-throughput characterization processes with high consistency, high efficiency, and low cost.

## B. Recycling

In the early history of ICs, damaged integrated devices were deemed unsuitable for reuse and were usually discarded. However, with the rapid advancement of fabrication techniques, modern recycling technologies have changed the way for handling discarded integrated devices. The recycling of these devices provides a sustainable alternative to their disposal with added benefits such as resource conservation, cost reduction, environmental impact reduction, and circular economy promotion. For bulk integrated photonic devices, the recycling process typically involves chemical etching, mechanical polishing, and hydrothermal treatments,[159-161] ensuring that the recycled devices meet the necessary standards for reuse in new fabrication. In hybrid integrated photonic devices incorporating 2D materials, the ultrathin 2D films tend to be more susceptible to damage compared to the bulk integrated devices. In most cases, damage is only limited to 2D materials themselves. Therefore, the removal and retransfer of 2D films are crucial for reusing these devices, particularly given that fabricating new integrated devices is usually more intricate and costly than recoating 2D films.

**Fig. 9** illustrates typical processes for recycling hybrid integrated photonic devices incorporating 2D materials. First, 2D materials are transferred and patterned on bare integrated photonic devices to fabricate a product for use. Next, the fabricated device experiences damage to the 2D materials due to factors such as exposure to high temperature, humid environment, bending or stretching, and exposure to high-power laser irradiation. Finally, the undamaged integrated photonic device is reused for a new product after removing the damaged 2D film and recoating a new one, which also represents the start of a new cycle.

Flat polymer films, such as PDMS or PMMA, have been employed for mechanically peeling off 2D materials from underlying substrates.[162] This technique leverages strong adhesion between the polymer and the 2D material, allowing for effective removal of 2D materials without introducing chemical contaminants and significant damage to the substrates. In addition, wet-chemical etching, which typically employs specific solvents like acetone, acid, and tetramethylammonium hydroxide, facilitates selective dissolution of certain 2D materials such as GO and TMDCs.[163] It allows for controlled removal of 2D films while preserving the integrity of neighboring structures. Oxygen plasma treatment via reactive ion bombardment can efficiently oxidize and etch away 2D films like GO, TMDCs and BP, and maintain minimal impact on the remaining components.[164] Laser ablation employs focused laser pulses to vaporize or break chemical bonds in target areas, offering a highly localized and non-contact method for the precise removal of 2D films with minimal thermal damages. It has been used to effectively remove graphene and hBN.[165]

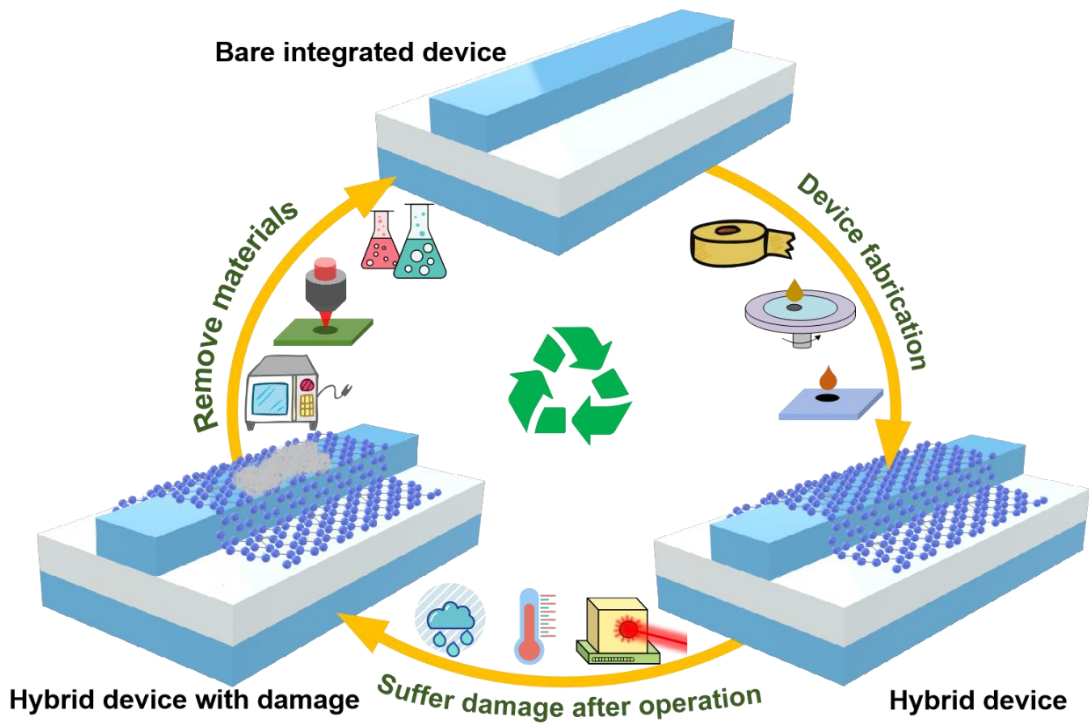

**FIG. 9.** Recycling of hybrid integrated photonic devices incorporating 2D materials.

To date, the recycling techniques mentioned above are mainly used in laboratory research, and there are several challenges associated with their implementation in sustainable industrial manufacturing processes. First, the aforementioned recycling methods are more widely employed for passive integrated photonic devices. For active integrated photonic devices such modulators, switches, and photodetectors, electrodes are typically placed on top of the 2D films. Consequently, recycling these devices involves more complex processes and incurs higher costs. In addition, the recycling costs should be weighed against production costs. Integrated devices are typically produced on a wafer scale, featuring multiple devices on a single wafer. Recoating a single compact device with 2D materials may involve more complex processes and operations compared to doing so on an entire wafer, particularly given that additional alignment tools are needed to ensure the precise placement of 2D films on the target integrated photonic devices. As a result, the recycling process can be more costly than producing a new device from a batch at the wafer scale. Driven by ongoing advancements in technology and growing awareness of sustainability, more devices recycling techniques are expected to develop such as modular designs and flexible disassembly,[166] leading to further reduced recycling costs.

## C. Service life

The service life is a crucial indicator of the quality of commercial products. In industrial production, improving fabrication technology to extend service life is critical for boosting product competitiveness and appeal. Conventional IC devices have different lifespans depending on their ambient conditions and using frequency, which typically range from 3 to 10 years.[167] Given the fact that 2D materials are typically more sensitive to environmental factors than bulk integrated devices, the service life of hybrid integrated devices incorporating 2D materials can be shortened.

To improve the service life of hybrid integrated devices, it is necessary to understand the degradation mechanisms of 2D materials and make targeted improvements. One of the dominant mechanisms is the presence of water and oxygen in the air. Taking BP as an example, BP flakes are highly hydrophilic, and long-term exposure to air leads to etching of the material.[168, 169] Testing results showed that degradation begins to occur on BP surface after about one hour of exposure to air.[170] In addition, theoretical calculations show that dipole-dipole interactions between water molecules and phosphorene can induce substantial lattice distortion, shrinking the lattice by ~25% and altering the band gap by ~20%.[169] Experimental demonstrations also show that few-layer BP flakes immersed in oxygen-enriched water were completely etched away,[171] reflecting the rapid degradation of BP caused by the combined effects of oxygen and water. Aside from BP, even highly stable 2D materials like graphene can experience performance degradation caused by water and oxygen, particularly for highly sensitive devices such as modulators and photodetectors.[29, 57, 172, 173]

Another important mechanism is light-induced oxidation, where incident light serves as an excitation source to trigger chemical reactions between the materials and oxygen. This process could generate reactive oxygen species that interact with 2D materials such as BP and TMDCs, resulting in structural alterations, loss of crystallinity, and diminished electronic or optical properties. Studies have shown that UV light is the main catalyst for this oxidation process, with the degree of oxidation increasing in proportion to the light intensity. Due to the quantum confinement effect, the degree of degradation is also influenced by the thicknesses of 2D films.[169]

The degradation mechanisms mentioned above can be enhanced by increasing temperatures, which accelerate chemical reactions that undermine material integrity and result in increased degradation rates. Studies have shown that rising temperatures can increase defect mobility and exacerbate thermal effects on the electrical and optical properties of many 2D materials. For instance, high temperatures in $MoS_2$ can cause sulfur vacancies that disrupt the crystal structure and alter the bandgap.[174] In addition, increasing temperatures in GO can break the chemical bonds between the oxygen-containing functional groups and the carbon network, resulting in the reduction of GO and changes in its properties.[123]

Some other degradation mechanisms can also affect the stability and performance of 2D materials in hybrid integrated devices. Mechanical stress, for instance, can introduce defects in the crystal lattice, leading to a decrease in electrical conductivity and optical performance. The stress can arise from external forces or thermal expansion mismatches between 2D films and their substrates, which could result in cracking or delamination. In addition, 2D films can easily adsorb dust and other airborne contaminants, which, when combined with other mechanisms such as temperature fluctuations and mechanical stress, create synergistic effects that further accelerate their degradation.

To avoid the degradation arising from the above factors, various encapsulation methods have been investigated, as we discussed in **Part D** of **Section III**. In recent years, growing attention has focused on extending the service life of integrated photonic devices incorporating 2D materials. For example, a graphene-BP photodetector with long-term stability was demonstrated in Ref. [175], where the robust graphene functioned as both an encapsulation layer and a highly efficient transport layer. Under ambient conditions, the device's photoresponsivity exhibited a minimal change of ~6% after 60 days of operation under consistent illumination. This represents a significant improvement in the service life compared to a pure BP device under similar conditions. In Ref. [176], CsPbI$_3$ perovskite solar cells with high stability were demonstrated, where a 2D Cs$_2$PbI$_2$Cl$_2$ capping layer was deposited between the perovskite active layer and the hole-transport layer. These devices did not degrade at 35°C and degraded by 20% of their initial efficiency for over 2100 hours at 110°C under constant illumination, significantly outperforming the degradation of 80% within few hundred to thousand hours reported for other perovskite solar cells. When operating continuously at 35°C, it is predicted that the lifetime of the encapsulated solar cells can reach 51,000 ± 7000 hours (*i.e.*, >5 years).

Besides improvements in device fabrication, it is crucial to regulate environmental conditions during storage and use to maximize the service life of related products. For instance, maintaining low humidity levels can help prevent degradation caused by moisture, and temperature control can minimize thermal effects and related damage. Furthermore, shielding devices from direct light exposure, especially UV light, can lower the risk of light-induced degradation, thereby extending the service life of hybrid integrated devices.

## D. Environmental implications

Another important issue to consider when developing commercial products is their environmental implications. From the start of production through to end-of-life disposal, commercial products have many potential environmental implications that can possibly cause air, soil, and water pollution, as well as potential health risks such as released toxic gases and remnant toxic waste.

For bulk integrated photonic devices, the CMOS fabrication processes can involve environmental implications. For example, photoresist and developing solution used for photolithography typically have high toxicity and corrosiveness, and improper handling or disposal of these chemicals can lead to contamination of soil and water.[177] In addition, the CMOS production lines are featured by high electrical energy consumption, and this energy-intensive nature leads to increased carbon emission that will exacerbate climate change.[178,179] Greenhouse gases, such as fluorinated compounds (*e.g.*, $SF_6$, $NF_3$, $CF_4$, $CHF_3$), are commonly used for dry etching, cleaning CVD chambers, and epitaxial material growth, which also have a negative environmental impact.[179] Along with the rapid progress in commercialization of related products and development of fabrication techniques, many environmental issues for the IC industry have been adequately considered and effectively mitigated in modern manufacturing facilities, steering the entire industry toward enhanced sustainability.

Compared with bulk integrated photonic devices, on-chip integration of 2D materials involves additional fabrication processes that may present new environmental challenges. For instance, the synthesis of TMDCs such as $MoS_2$ often involves toxic precursors like hydrogen sulfide ($H_2S$) or selenium (Se), which can pose health risks and environmental hazards.[180] The wet-transfer method involves various solvents and chemicals, such as isopropanol and acid, which are harmful if directly released into the environment.[72] The reaction process in CVD can produce waste gases that may contain unreacted precursors such as metalorganic compounds, silanes, as well as toxic byproducts like hydrogen chloride and volatile organic compounds, which can also contaminate air and water.[181] The device packaging often involves some non-biodegradable materials such as PMMA and PDMS, which may lead to waste disposal issues.[137,138]

In addition to environmental pollution generated during the fabrication process, 2D materials themselves can exhibit toxicity and pose health risks to humans. Small 2D material flakes can easily be transported and accumulated in cells or respiratory tract after inhalation.[182] Studies have shown that the in vivo toxicity of 2D materials is significantly influenced by the preparation methods, along with factors such as flake size, surface area, thickness, and the amount and type of oxygen functionalization groups. It was also found that the toxicity of TMDCs was lower than of GO, and BP exhibited toxicity levels that were intermediate between GO and TMDCs.[183] This work will benefit all microcomb based applications[184–207], particularly microwave photonics[208-308], quantum optical photonics[309-325] as well as potentially being able to exploit novel 2D nonlinear materials [326–358] and structures[359–367].

To date, research on the environmental implications of hybrid integrated devices incorporating 2D materials remains relatively limited. With continuous advancements in fabrication techniques and commercialization, it is expected that increasing focus will be placed on the environmental implications of relevant products. In the future, more environment-friendly materials and fabrication methods will be adopted to replace those that are prone to generating toxic byproducts. The shift towards sustainable solutions will not only benefit the environment but also align with consumer demands for greener products, which will ultimately foster a more responsible and sustainable industry for developing relevant products.

## V. CONCLUSION

Facilitated by the superior properties of 2D materials and the rapid progress in related fabrication techniques, hybrid integrated photonic devices incorporating these promising materials are rapidly advancing towards industrial manufacturing and commercialization. In this paper, we provide a perspective on the developments of this field. We first provide an overview of recent progress towards commercialization. Next, we summarize state-of-the-art fabrication techniques, and highlight both their advantages and limitations in relation to industrial manufacturing. Finally, we address key considerations for commercialization, including fabrication standards, recycling, service life, and environmental implications. In the future, we believe that the enhanced collaboration between academia and industry will further liberate research on 2D material integrated photonics from lab, leading to more and more reliable commercial products that are easily accessible to end users.

## Conflict of Interest

The authors have no conflicts to disclose.